\documentclass{elsart}
\usepackage{graphics}
\usepackage{bm}

\begin{document}

\begin{frontmatter} 

\title{Neutrino effects in two-body electron-capture measurements at GSI} 

\author{Avraham Gal}
\ead{avragal@vms.huji.ac.il}
\address{Racah Institute of Physics, The Hebrew University, Jerusalem 
91904, Israel} 

\date{\today} 

\begin{abstract} 
I conjecture that the time modulated decay rates reported in single ion 
measurements of two body electron capture decay of hydrogen like heavy 
ions at GSI may be related to neutrino spin precession in the static 
magnetic field of the storage ring. These `GSI Oscillations' arise from 
interference between amplitudes of decay within and without the magnetic 
field, a scenario that requires a Dirac neutrino magnetic moment six times 
lower than the Borexino solar neutrino upper limit of 0.54 x 10E(-10) Bohr 
magneton. I also show in a way not discussed before that the time modulation 
associated with interference between massive neutrino amplitudes, if such 
interference could arise, is of a period at least four orders of magnitude 
shorter than reported and must average to zero given the time resolution 
of the GSI measurements. 
\end{abstract}


\end{frontmatter}

\section{Introduction} 
\label{sec:intro} 

Measurements of weak interaction decay of multiply ionized heavy ions 
coasting in the ion storage-cooler ring ESR at the GSI laboratory, 
since the first report in 1992 \cite{GBB92}, open up new vistas for 
dedicated studies of weak interactions. In particular, electron 
capture (EC) decay rates in hydrogen-like and helium-like $^{140}$Pr 
ions have been recently measured for the first time \cite{LBG07} by 
following the motion of the decay ions (D) and the recoil ions (R). 
The overall decay rates $\lambda_{\rm EC}$ of these two-body 
$^{140}{\rm Pr} \to {^{140}{\rm Ce}} + \nu$ EC decays, in which no neutrino 
$\nu$ is detected, are well understood within standard weak interaction 
calculations of the underlying $e^-p\to\nu_e n$ reaction \cite{PKB08,IFR07}. 
However, a time-resolved decay spectroscopy applied subsequently to the 
two-body EC decay of H-like $^{140}$Pr and $^{142}$Pm single ions revealed 
an oscillatory behavior, or more specifically a time modulation of the 
two-body EC decay rate \cite{LBW08}: 
\begin{equation} 
\lambda_{\rm EC}(t)=\lambda_{\rm EC}[1+a_{\rm EC}\cos(\omega_{\rm EC} t + 
\phi_{\rm EC})], 
\label{eq:omega} 
\end{equation} 
with amplitude $a_{\rm EC} \approx 0.2$, and angular frequency 
$\omega^{\rm lab}_{\rm EC} \approx 0.89~{\rm s}^{-1}$ 
(period $T^{\rm lab}_{\rm EC} \approx 7.1$~s) in the 
laboratory system which is equivalent in the rest frame of the decay ion 
to a minute energy $\hbar \omega_{\rm EC} \approx 0.84 \times 10^{-15}$~eV. 
Subsequent experiments on EC decays of neutral atoms in solid environment 
have found no evidence for oscillations with periodicities of this order of 
magnitude \cite{VCD08,FBH08}. Thus, the oscillations observed in the GSI 
experiment could have their origin in some characteristics of the H-like ions, 
produced and isolated in the ESR, and in the electromagnetic fields specific 
to the ESR which are not operative in normal laboratory experiments. It is 
suggested here, in Sect.~\ref{sec:magfield}, that the `GSI Oscillations' could 
indeed be due to the magnetic field which stabilizes and navigates the motion 
of the ions in the ESR. 

Several works, by Kienle and collaborators, relegated the `GSI Oscillations' 
to interference between neutrino mass eigenstates that evolve coherently 
from the electron neutrino $\nu_e$ \cite{IRK08,Fab08,KKi08,IKP08,IK09}. 
This idea apparently also motivated the GSI experiment \cite{LBW08}. 
Such interferences, according to these works, lead to oscillatory behavior 
given by Eq.~(\ref{eq:omega}) with angular frequency $\omega_{\nu_e}$ 
where, again in the decay-ion rest frame, 
\begin{equation}  
\hbar\omega_{\nu_e}=\frac{\Delta (m_{\nu}c^2)^2}{2M_Dc^2}\approx 
0.29 \times 10^{-15}~{\rm eV}. 
\label{eq:delE} 
\end{equation}  
Here, $\Delta (m_{\nu}c^2)^2 = (0.76 \pm 0.02) \times 10^{-4}$~eV$^2$ from 
accumulated solar $\nu$ plus KamLAND reactor $\bar{\nu}$ data \cite{SNO08} 
for the two mass-eigenstate neutrinos that almost exhaust the coupling 
to $\nu_e$, and $M_D\approx 130$~GeV/$c^2$ is the mass of the decay ion 
$^{140}{\rm Pr}^{58+}$. Although the value of $\hbar\omega_{\nu_e}$ on the 
r.h.s. of Eq.~(\ref{eq:delE}) is about three times smaller than the value of 
$\hbar \omega_{\rm EC}$ required to resolve the `GSI Oscillation' puzzle, 
getting down to this order of magnitude nevertheless presents a remarkable 
achievement if correct.{\footnote{Eq.~(\ref{eq:delE}) was also obtained by 
Lipkin \cite{Lip08} assuming interference between two unspecified components 
of the initial state with different momenta and energies that can both decay 
into the same final state, an electron neutrino and a recoil ion with definite 
energy and momentum. This scenario was criticized by Peshkin \cite{Pes08}.}} 
However, it is shown here in Sect.~\ref{sec:osc} by following the methodology 
of Ref.~\cite{IK09} that the correct energy scale under circumstances allowing 
oscillatory behavior is given by 
\begin{equation} 
\hbar\Omega_{\nu_e}=\frac{\Delta (m_{\nu}c^2)^2}{2E_{\nu}}\approx 0.95 
\times 10^{-11}~{\rm eV}, 
\label{eq:DelE} 
\end{equation} 
where $E_{\nu}\approx 4$~MeV is a representative value for neutrino 
energy in the H-like $^{140}{\rm Pr}\to{^{140}{\rm Ce}}+\nu_e$ and 
$^{142}{\rm Pm}\to {^{142}{\rm Nd}}+\nu_e$ EC decays \cite{LBW08}. 
The energy $\hbar\Omega_{\nu_e}$ is larger by over four orders of 
magnitude than $\hbar\omega_{\rm EC}$ or $\hbar\omega_{\nu_e}$ given 
by Eq.~(\ref{eq:delE}), and so it would lead to modulation period 
shorter by over four orders of magnitude than the 7~s period reported 
by the GSI experiment. Given a time measurement resolution of order 
$0.5$~s \cite{LBW08}, the effect of such oscillatory behavior would 
average out to zero. 

Other authors \cite{Giu08,BLS08,KKL08,CGL09,Mer09,Fla09} have 
rejected any link between neutrino mass eigenstates and the EC decay 
rate oscillatory behavior reported by the GSI experiment \cite{LBW08}, 
the underlying argument being that since no neutrino is detected, 
the EC decay rate sums incoherently over neutrino mass eigenstates, 
whereas any oscillatory behavior requires interference between amplitudes 
summed upon coherently. It is instructive, however, to demonstrate this 
assertion also by adopting the methodology of Ref.~\cite{IK09}, but with 
a caveat explained below. To this end the time-dependent EC transition 
amplitude $A_{\nu_e}(i \to f;~t)$, from initial state $i$ (D injected at 
time $t=0$) to a final state $f$ (R plus a coherent combination of neutrino 
mass eigenstates at time $t$), is written in terms of transition amplitudes 
$A_{\nu_j}(i\to f;~t)$ that involve propagating mass-eigenstates neutrinos 
$\nu_j$ as 
\begin{equation} 
A_{\nu_e}(i\to f;~t) = \sum_jU_{ej}A_{\nu_j}(i\to f;~t), 
\label{eq:Ae} 
\end{equation} 
where $U_{e j}$ is a neutrino mixing matrix element of the $3 \times 3$ 
unitary matrix $U$ 
\begin{equation} 
|\nu_\alpha\rangle = \sum_{j=1}^3 U_{\alpha j}^* \, |\nu_j\rangle~~~~
(\alpha = e, \mu, \tau) 
\label{eq:U} 
\end{equation} 
between the emitted electron-neutrino $\nu_e$ and a mass-eigenstate neutrino 
$\nu_j$ \cite{ADA08}. For times of order seconds, appropriate to the `GSI 
Oscillations', the coherence implied by Eq.~(\ref{eq:Ae}) is still in 
effect \cite{Mer09} and the flavor basis is of physical significance. 
If the GSI experiment were to detect neutrino $\nu_{\beta}$ by a flavor 
measurement, the corresponding amplitude would have been generated by 
projecting Eq.~(\ref{eq:Ae}) onto flavor $\beta$: 
\begin{equation} 
A_{\nu_e \to \nu_{\beta}}(i\to f;~t) = \sum_jU_{ej}A_{\nu_j}(i\to f;~t)
U^*_{\beta j}, 
\label{eq:Abeta} 
\end{equation} 
in close analogy with the discussion of neutrino oscillations in dedicated
oscillation experiments (Eq.~(13.4) in Ref.~\cite{ADA08}). The probability 
associated with the amplitude (\ref{eq:Abeta}) is then given by 
\begin{equation} 
{\cal P}_{\nu_e \to \nu_{\beta}}(i\to f;~t)=|A_{\nu_e \to \nu_{\beta}}
(i\to f;~t)|^2. 
\label{eq:Pe} 
\end{equation} 
Interference terms $A_{\nu_j}A^*_{\nu_{j'}}$ will arise in 
${\cal P}_{\nu_e \to \nu_{\beta}}(i\to f;~t)$, leading to oscillations as 
shown in Sect.~\ref{sec:osc}. Since the GSI experiment does not detect any 
neutrino, the overall probability is the sum of probabilities 
${\cal P}_{\nu_e \to \nu_{\beta}}(i\to f;~t)$ over {\it all} three flavors 
$\beta$. The probability of observing the transition $i\to f$, in which D 
decays to R and a neutrino is emitted but remains undetected, is thus given 
by 
\begin{equation} 
{\cal P}_{\nu_e}(i\to f;~t) =\sum_{\beta}|\sum_j U_{ej}A_{\nu_j}(i\to f;~t)
U^*_{\beta j}|^2. 
\label{eq:interf} 
\end{equation} 
Note that the probability ${\cal P}_{\nu_e}(i\to f;~t)$ cannot be written as 
a square of {\it one} amplitude, simply because it involves different-flavor 
final states which require measurement schemes differing from each other and 
therefore adding up incoherently. Using the unitarity of the mixing matrix 
$U$, the summation over $\beta$ in Eq.~(\ref{eq:interf}) gets rid of the 
interference terms, leading to the final expression: 
\begin{equation} 
{\cal P}_{\nu_e}(i\to f;~t)=\sum_{j}|U_{ej}|^2|A_{\nu_j}(i\to f;~t)|^2
\approx |A_{\nu}(i\to f;~t)|^2, 
\label{eq:incoherent} 
\end{equation}  
where the dependence of the absolute-squared terms 
$|A_{\nu_j}(i\to f;~t)|^2$ on the species $\nu_j$ was neglected,{\footnote
{This neglect does not hold for interference terms $A_{\nu_j}A^*_{\nu_{j'}}$, 
$j \neq j'$, which give rise to oscillatory behavior, as discussed in 
Sect.~\ref{sec:osc}.}} enabling repeated use of unitarity. The final result, 
Eq.~(\ref{eq:incoherent}), is that the probability 
${\cal P}_{\nu_e}(i\to f;~t)$ for the two-body EC decay to occur is what 
standard weak interaction theory yields for a massless electron neutrino, 
regardless of its coupling to the mass-eigenstate neutrinos. This holds true 
also for the total EC decay rate which is obtained by time differentiation 
of ${\cal P}_{\nu_e}(i\to f;~t)$ and which is found identical with the 
time-independent decay rate $\lambda_{\rm EC}$ derived ignoring neutrino 
mixing. Thus, although the mass-eigenstate components of the emitted 
neutrino oscillate against each other, the total EC decay rate does not 
exhibit any oscillatory behavior owing to the unitarity of the matrix $U$, 
Eq.~(\ref{eq:U}), which transforms incoherence in one basis into incoherence 
in the other basis. If $U$ is nonunitary, the above argumentation breaks 
down, but this does not spoil the more straightforward argumentation that 
mass-eigenstate neutrinos, as distinct mass particles, have to be summed 
upon incoherently; one then goes directly from the amplitude $A_{\nu_e}$, 
Eq.~(\ref{eq:Ae}), into the probability ${\cal P}_{\nu_e}$, 
Eq.~(\ref{eq:incoherent}), which indeed is an incoherent sum over the 
neutrino mass-eigenstates $\nu_j$.

\section{Detection of a flavor neutrino, neutrino oscillations} 
\label{sec:osc} 

Oscillatory behavior of EC decay rates is possible when a neutrino of 
a given flavor is detected. The relatively small energy of order few MeV 
released in EC limits the detected neutrino to $\nu_e$. Here I show within 
a straightforward `gedanken' extension of the GSI experiment, in which 
an electron-neutrino $\nu_e$ is detected, that the corresponding angular 
frequency of the oscillations is given by $\hbar\Omega_{\nu_e}$, 
Eq.~(\ref{eq:DelE}). To this end, the specific time-dependent first-order 
perturbation theory amplitudes $A_{\nu_j}(i\to f;~t)$ introduced by Ivanov 
and Kienle \cite{IK09} are followed as much as possible: 
\begin{equation} 
A_{\nu_j}(i\to f;~t) = -i\int^t_0\langle 
f(\vec{q}\,)\nu_j(\vec{k}_j)|H_{e\nu_j}(\tau)|i(\vec{0}\,)\rangle d\tau, 
\label{eq:Aej} 
\end{equation} 
with a weak-interaction Hamiltonian for the leptonic transition  
$e^-\to\nu_j$ given by 
\begin{equation}  
{\cal H}_{e\nu_j}(\tau)=\frac{G_F}{\sqrt{2}}V_{ud}\int{d^3x
[\bar{\psi}_n\gamma^{\lambda}(1 - g_A\gamma^5) \psi_p]
[\bar{\psi}_{\nu_j}\gamma_{\lambda}(1 - \gamma^5)\psi_{e^-}]}. 
\label{eq:Hej} 
\end{equation} 
Here, $x=(\tau,\vec{x}\,)$, $G_F$ is the Fermi constant, $V_{ud}$ is the CKM 
matrix element, $g_A$ is the axial coupling constant, and with $\psi_n(x)$, 
$\psi_p(x)$, $\psi_{\nu_j}(x)$ and $\psi_{e^-}(x)$ denoting neutron, proton, 
mass-eigenstate neutrino $\nu_j$ and electron field operators, respectively. 
EC decays occur at any time $\tau$ within [$0,t$], from time $t'=0$ of 
injection of D into the ESR to time $t'=t$ of order seconds and longer at 
which the EC decay rate is evaluated. In the single-ion GSI experiment 
\cite{LBW08} the heavy ions revolve in the ESR with a period of order 
$10^{-6}$~s and their motion is monitored nondestructively once per 
revolution. The decay is defined experimentally by the {\it correlated} 
disappearance of D and appearance of R, but the appearance in the frequency 
spectrum is delayed by times of order 1~s needed to cool R. The order of 
magnitude of the experimental time resolution is similar, about 0.5~s, 
as reflected in the time intervals used to exhibit the experimental decay 
rates ${\cal R}(t)$ in Figs.~3,4,5 of Ref.~\cite{LBW08}. 
The decay rates determined in the ESR appear to agree with those measured 
elswhere, e.g. for $^{142}$Pm \cite{VCD08}, and this consistency suggests 
that details of kinematics and motion of the heavy ions in the storage ring 
affect little the overall decay rates which are evaluated here in conventional 
time-dependent perturbation theory. Therefore, it is plausible to assume that 
the evolution of the final state in these single-ion EC measurements at GSI 
proceeds over times of order 1~s which is used here as a working hypothesis. 

To obtain the time dependence of the amplitude $A_{\nu_j}(i \to f;~t)$ 
(similarly structured to Eq.~(6) of Ref.~\cite{IK09}), recall that the 
time dependence of the integrand in Eq.~(\ref{eq:Aej}) is given by 
$\exp({\rm i}\Delta_j\tau)$ where{\footnote{From here on $\hbar=c=1$ units 
are almost exclusively used.}} 
\begin{equation} 
\Delta_j(\vec{q}\,) = E_R(-\vec{q}\,) + E_j(\vec{q}\,) - M_D 
\label{eq:DeltaE} 
\end{equation} 
with 
\begin{equation} 
E_R=\sqrt{M_R^2+(-\vec{q}\,)^2}, \,\,\,\,\,\,
E_j=\sqrt{m_j^2+\vec{q}\,^2} 
\label{eq:E_Rj} 
\end{equation} 
for the recoil ion and neutrino $\nu_j$ energies, respectively, 
in the decay-ion rest frame. Integrating on this time dependence 
results in a standard time-dependent perturbation-theory energy-time 
dependence \cite{Baym69} 
\begin{equation} 
A_{\nu_j}(i \to f;~t) \sim \frac{1 - \exp({\rm i}\Delta_j t)}{\Delta_j}. 
\label{eq:TD1} 
\end{equation} 
Finally, the EC partial decay rate ${\cal R}_{\nu_e \to \nu_e}(i\to f;~t)$ 
is obtained from the probability ${\cal P}_{\nu_e \to \nu_e}(i\to f;~t)$, 
Eq.~(\ref{eq:Pe}), by differentiating: ${\cal R}={\partial}_t{\cal P}$. 
The term `partial' applied to the rate ${\cal R}_{\nu_e \to \nu_e}$ owes to 
its limitation to the detection of one particular kind of flavor neutrinos: 
depleted electron neutrinos. Using Eq.~(\ref{eq:TD1}) for the time dependence 
of $A_{\nu_j}(i \to f;~t)$, one gets for the contribution of any $j'=j$ non 
oscillatory term to ${\cal R}_{\nu_e \to \nu_e}$: 
\begin{equation} 
{\cal R}_{\nu_j} = \frac{d}{dt}|A_{\nu_j}(i \to f;~t)|^2 \sim 
\frac{2\sin (\Delta_j t)}{\Delta_j} \to 2\pi\delta(\Delta_j), 
\label{eq:TD2} 
\end{equation} 
where the last step requires a sufficiently long time $t$. 
The properly normalized contribution of these terms to 
${\cal R}_{\nu_e \to \nu_e}(i\to f;~t)$ is given by 
\begin{equation} 
\sum_j {\cal R}_{\nu_j} = \lambda_{\rm EC}\sum_j |U_{ej}|^4 \delta(\Delta_j). 
\label{eq:jj} 
\end{equation} 
Similarly, the contribution of the $j' \neq j$ oscillatory terms to the 
EC partial decay rate ${\cal R}_{\nu_e \to \nu_e}(i\to f;~t)$, again for 
sufficiently long times, is given by 
\begin{equation} 
\lambda_{\rm EC}\sum_{j>j'} |U_{ej}|^2 |U_{ej'}|^2 [\delta(\Delta_j)+
\delta(\Delta_{j'})] \cos[(\Delta_j - \Delta_{j'})t]. 
\label{eq:ivanov} 
\end{equation} 
The $\delta$ symbols in Eqs.~(\ref{eq:jj}) and (\ref{eq:ivanov}) differ from 
Dirac $\delta$ functions in that no further integration on the implied c.m. 
momentum $\vec q$ has to be done. Their meaning is straightforward for the 
non oscillatory terms, but more delicate for the oscillatory terms in which 
the sum of $\delta$ symbols imply that $\Delta_j - \Delta_{j'}$ be 
evaluated for momentum once derived from the constraint 
$\Delta_j(\vec{q}\,)=0$ and once derived from $\Delta_{j'}(\vec{q}\,)=0$. 
On each occasion, using a generic notation $k$ for the momentum implied by 
each one of the $\delta$ symbols, one obtains to an excellent approximation 
\begin{equation} 
\Delta_j(k)-\Delta_{j'}(k)= E_j(k)-E_{j'}(k) = \hbar \Omega_{jj'}, 
\label{eq:correct} 
\end{equation} 
where $\Omega_{jj'}$ is related to $\Omega_{\nu_e}$ of 
Eq.~(\ref{eq:DelE}): 
\begin{equation} 
\hbar \Omega_{jj'} = \frac{m_j^2-m_{j'}^2}{2E_{\nu}} 
\approx \hbar \Omega_{\nu_e}. 
\label{eq:Omegajj'} 
\end{equation} 
The requirement of 
{\it sufficiently long times} for Eq.~(\ref{eq:ivanov}) to hold translates in 
the present case to requiring $t\gg\Omega_{\nu_e}^{-1}\sim 7\times 10^{-5}$~s, 
which is comfortably satisfied given the experimental time resolution scale of 
$\sim 0.5$~s \cite{LBW08}. 

The final expression for the depleted $\nu_e$ rate is obtained by integrating 
over the $\delta$ symbols in Eqs.~(\ref{eq:jj}) and (\ref{eq:ivanov}), 
resulting in 
\begin{equation} 
{\cal R}_{\nu_e \to \nu_e}(i\to f;~t)=\lambda_{\rm EC}\{\sum_j 
|U_{ej}|^4 + 2\sum_{j>j'} |U_{ej}|^2 |U_{ej'}|^2 \cos (\Omega_{jj'}t)\}. 
\label{eq:our} 
\end{equation} 
Using the unitarity of $U$, Eq.~(\ref{eq:our}) may be simplified 
to the following form: 
\begin{equation} 
{\cal R}_{\nu_e \to \nu_e}(i\to f;~t)=\lambda_{\rm EC}\{1-4\sum_{j>j'}
|U_{ej}|^2 |U_{ej'}|^2~{\sin}^2(\frac{\Omega_{jj'}}{2}t)\}. 
\label{eq:Kayser1} 
\end{equation}    
This expression is identical with the probability for $\nu_e\to\nu_e$ 
oscillation in neutrino spatial oscillation experiments (Eq.~(13.9) in 
Ref.~\cite{ADA08}) upon making the identification $t=L/c$, where $L$ is 
the distance traversed by the neutrino between its source and the detector. 
A more rigorous wave-packet treatment is required to justify this transition 
from $t$ to $L$ \cite{AS09}. A further simplification of 
Eq.~(\ref{eq:Kayser1}) occurs when only two of the mass-eigenstate neutrinos 
are coupled to $\nu_e$: 
\begin{equation} 
{\cal R}_{\nu_e \to \nu_e}(i\to f;~t)=\lambda_{\rm EC}\{1-{\sin}^2 2\theta~ 
{\sin}^2(\frac{\Omega_{\nu_e}}{2}t)\}, 
\label{eq:Kayser2} 
\end{equation} 
where $\theta$ is the $\nu_1 \leftrightarrow \nu_2$ mixing angle 
(cf. Eq.~(13.20) in Ref.~\cite{ADA08}). 
Note that it is the neutrino energy $E_{\nu}$ to which the period of 
oscillations is proportional, not to the mass $M_D$ of the decaying heavy 
ion in the GSI experiments.{\footnote{Ivanov and Kienle \cite{IK09} overlooked 
this distinction by using in Eq.~(\ref{eq:correct}) simultaneously {\it on 
energy shell} momentum values $k_j$ and $k_{j'}$ implied by $\delta(\Delta_j)$ 
and $\delta(\Delta_{j'})$ respectively, and replacing $\Delta_j - \Delta_{j'}$ 
in the oscillatory terms of Eq.~(\ref{eq:ivanov}) by $E_j(k_j)-E_{j'}(k_{j'}) 
\approx \hbar\omega_{\nu_e}$, Eq.~(\ref{eq:delE}). A similar error was made by 
Kleinert and Kienle when evaluating Eq.~(54) in Ref.~\cite{KKi08}.}}

\section{Magnetic field effects} 
\label{sec:magfield}

The preceding discussion ignored a possible role of the electromagnetic 
fields surrounding the ESR for guidance and stabilization of the 
heavy-ion motion. The nuclei $^{140}$Pr and $^{142}$Pm in the GSI 
experiment \cite{LBW08} have spin-parity $I_i^{\pi}=1^+$, and the 
electron-nucleus hyperfine interaction in the decay ion forms a doublet 
of levels $F_i^{\pi}=({\frac{1}{2}}^+,{\frac{3}{2}}^+)$, 
the `sterile' ${\frac{3}{2}}^+$ level lying about 1 eV above the `active' 
${\frac{1}{2}}^+$ g.s. from which EC occurs to a $F_f=\frac{1}{2}$ final 
state of a fully ionized recoil ion with spin-parity $I_f^{\pi}=0^+$ plus 
a left-handed neutrino of spin $\frac{1}{2}$.{\footnote{The subscript $f$ 
in this section relates to both the recoil ion and the neutrino.}} The 
lifetime of the $F_i^{\pi}={\frac{3}{2}}^+$ excited level is of order 
$10^{-2}$~s, so that it de-excites sufficiently rapidly to the 
$F_i^{\pi}={\frac{1}{2}}^+$ g.s. \cite{LBG07,IFR07}. 
Periodic excitations of this `sterile' state cannot explain the reported 
time dependence and intensity pattern \cite{WSI09}. 
The static magnetic field which is perpendicular to the ESR, $B=1.19$~T 
for $^{140}{\rm Pr}$ \cite{Fas09private}, gives rise to precession of the 
$F_i^{\pi}={\frac{1}{2}}^+$ initial-state spin with angular frequency 
$\omega_i$ of order $\hbar\omega_i \sim {\mu_B}B\approx 0.7\times 10^{-4}$~eV 
\cite{FIK09}, where $\mu_B$ is the Bohr magneton. The corresponding time scale 
of order $10^{-11}$~s is substantially shorter than even the ESR revolution 
period $t_{\rm revol} \approx 0.5\times 10^{-6}$~s, so any oscillation arising 
from this initial-state precession would average out to zero over 1 cm of the 
approximately 100 m long circumference. A nonstatic magnetic field could lead 
through its high harmonics to oscillations with the desired frequency between 
the magnetic substates of the $F_i^{\pi}={\frac{1}{2}}^+$ g.s. \cite{Pav10}, 
but the modulation amplitude $a_{\rm EC}$ expected for such harmonics is below 
a $1\%$ level, and hence negligible. Furthermore, the associated mixing 
between the two hyperfine levels $F_i^{\pi}=({\frac{1}{2}}^+,{\frac{3}{2}}^+)$ 
is negligible. In conclusion, no initial-state coherence effects are expected 
from internal or external electromagnetic fields in the GSI experiment. 

In the final configuration, interferences may arise from the precession of 
the neutrino spin in the static magnetic field of the ESR. The corresponding 
angular frequency $\omega_{\mu_{\nu}}$ is given by $\hbar\omega_{\mu_{\nu}} = 
{\mu_{\nu}}\gamma B<0.5\times 10^{-14}$~eV in the decay ion rest frame, due to 
the neutrino anomalous magnetic moment $\mu_{\nu}$ interacting with the static 
magnetic field $B$. Here, $\gamma=1.43$ is the Lorentz factor relating the 
rest frame to the laboratory frame, and $\mu_{\nu} < 0.54\times 10^{-10}\mu_B$ 
from the Borexino solar neutrino data \cite{Borexino08}. Below I show how the 
total EC rate gets time-modulated with angular frequency $\omega_{\mu_{\nu}}$. 
To agree with the reported GSI measurements, $\omega_{\mu_{\nu}}=
\omega_{\rm EC}$, a value of the electron-neutrino magnetic moment 
$\mu_{\nu} \sim 0.9 \times 10^{-11} \mu_B$ is required which is six times 
smaller than provided by the published Borexino solar neutrino upper limit 
\cite{Borexino08}.

\subsection{Interference due to a Dirac neutrino magnetic moment} 
\label{subsec:dirac}

For definiteness I first assume that neutrinos are Dirac fermions with only 
diagonal magnetic moments $\mu_{jk}=\mu_j \delta_{jk}$, and that these 
diagonal moments are the same for all 3 species: $\mu_j=\mu_{\nu}$. The 
emitted electron-neutrino is a left-handed lepton. The amplitude for producing 
it right-handed, namely with a positive helicity is negligible, of order 
$m_{\nu}/E_{\nu} < 10^{-7}$ and thus may be safely ignored. A static magnetic 
field perpendicular to the ESR flips the neutrino spin. Each of the 
mass-eigenstate components of the emitted neutrino will then precess, 
with amplitude $\cos (\omega_{\mu_{\nu}}t)$ for the depleted left handed 
components and with amplitude ${\rm i}\sin (\omega_{\mu_{\nu}}t)$ for the 
spin-flip right handed components \cite{FSh80}. Both are legitimate neutrino 
final states which are summed upon {\it incoherently}. 
The summed probability is of course time independent: 
$\cos^2(\omega_{\mu_{\nu}}t)+\sin^2 (\omega_{\mu_{\nu}}t)=1$. 
However, the magnetic field dipoles of the storage ring do not cover its full 
circumference, except for about $35\%$ of it \cite{Fas09private}. This results 
in interference between the decay amplitude $A^0_{\nu_j}$ for events with no 
magnetic interaction and the decay amplitude $A^{\rm m}_{\nu_j}$ for events 
undergoing magnetic interaction (superscript m) with depleted left handed 
components, i.e. with a superimposed amplitude of $\cos(\omega_{\mu_{\nu}}t)$: 
\begin{equation} 
A^0_{\nu_j} \sim -{\rm i}\int^t_0 \exp({\rm i}\Delta_j\tau) d\tau, \,\,\,\, 
A^{\rm m}_{\nu_j} \sim -{\rm i}\int^t_0 \exp({\rm i}\Delta_j\tau) 
\cos [\omega_{\mu_{\nu}}(t-\tau)] d\tau,   
\label{eq:Aj} 
\end{equation} 
using the same normalization as in Eq.~(\ref{eq:TD1}) for any of the 
left-handed mass-eigenstate neutrinos. This expression for $A^{\rm m}_{\nu_j}$ 
is a crude approximation, but has the merit of representing physically 
the sequential time dependence anticipated for magnetic interactions. 
For completeness, I also list the amplitude $A^{\rm R}_{\nu_j}$ for events 
undergoing magnetic interaction which have resulted in a right-handed neutrino 
(superscript R), with a superimposed amplitude of 
${\rm i}\sin (\omega_{\mu_{\nu}}t)$: 
\begin{equation} 
A^{\rm R}_{\nu_j} \sim -{\rm i}\int^t_0 \exp({\rm i}\Delta_j\tau) 
{\rm i}\sin [\omega_{\mu_{\nu}}(t-\tau)] d\tau. 
\label{eq:AjR} 
\end{equation} 
Repeating the same steps in going from amplitudes $A_{\nu_j}$, 
Eq.~(\ref{eq:TD1}), to decay rates ${\cal R}_{\nu_j}$, Eq.~(\ref{eq:TD2}), 
and adopting the same normalization, the decay rates associated with each 
one of these three amplitudes are given by: 
\begin{equation} 
{\cal R}^0_{\nu_j} \sim 2\pi\delta(\Delta_j), 
\label{eq:rate0j} 
\end{equation} 
\begin{equation} 
{\cal R}^{\rm m}_{\nu_j}\sim\frac{\pi}{2}[\delta(\Delta_j+\omega_{\mu_{\nu}}) 
+ \delta(\Delta_j-\omega_{\mu_{\nu}})](1+\cos(2\omega_{\mu_{\nu}}t)), 
\label{eq:ratemj} 
\end{equation} 
\begin{equation} 
{\cal R}^{\rm R}_{\nu_j}\sim\frac{\pi}{2}[\delta(\Delta_j+\omega_{\mu_{\nu}}) 
+ \delta(\Delta_j-\omega_{\mu_{\nu}})](1-\cos(2\omega_{\mu_{\nu}}t)). 
\label{eq:rateRj} 
\end{equation} 
Note that although the two latter expressions, for rates associated with the 
magnetic interaction, are time dependent, their sum is time independent as 
expected from summing incoherently over the two separate helicities. The only 
time dependence in this schematic model arises from interference of the two 
amplitudes $A^0_{\nu_j}$ and $A^{\rm m}_{\nu_j}$ for a left-handed neutrino. 
The sum of these partial rates, all of which correspond to $\nu_j$, 
and incorporating this interference, is given by 
\begin{eqnarray} 
{\cal R}_{\nu_j} \sim & |a_0|^2 2\pi\delta(\Delta_j) + |a_{\rm m}|^2 \pi 
[\delta(\Delta_j+\omega_{\mu_{\nu}}) + \delta(\Delta_j-\omega_{\mu_{\nu}})] 
\nonumber \\ 
+ & 2{\rm Re}(a_0 a^*_{\rm m}) \frac{\pi}{2}
[\delta(\Delta_j+\omega_{\mu_{\nu}}) + 2\delta(\Delta_j) 
+ \delta(\Delta_j-\omega_{\mu_{\nu}})] \cos (\omega_{\mu_{\nu}}t) 
\nonumber \\ 
& - 2{\rm Im}(a_0 a^*_{\rm m}) \frac{\pi}{2}
[\delta(\Delta_j+\omega_{\mu_{\nu}}) - \delta(\Delta_j-\omega_{\mu_{\nu}})] 
\sin (\omega_{\mu_{\nu}}t), 
\label{eq:ratej} 
\end{eqnarray} 
where $|a_{\rm m}|^2 \sim 0.35$ and $|a_0|^2 \sim 0.65$, with unknown relative 
phase between the probability amplitudes $a_{\rm m}$ and $a_0$ for undergoing 
or not undergoing magnetic interaction, respectively. 
Working out the complete normalization of this expression, the final rate 
expression is given by 
\begin{equation} 
{\cal R}_{\nu_e} = \lambda_{\rm EC}[1+2{\rm Re}(a_0 a^*_{\rm m}) 
\cos (\omega_{\mu_{\nu}}t)],  
\label{eq:ratenu} 
\end{equation} 
showing explicitly a time modulation of the kind Eq.~(\ref{eq:omega}) reported 
by the GSI experiment \cite{LBW08}. It is beyond the present schematic model 
to explain the magnitude of the modulation amplitude $a_{\rm EC}$ and the 
phase shift $\phi_{\rm EC}$, except that $|a_{\rm EC}|<1$. In particular, 
a more realistic calculation is required in order to study effects of 
departures from the idealized kinematics implicitly considered above by which 
{\it both} the recoil ion and the neutrino go forward with respect to the 
decay-ion instantaneous laboratory forward direction. Whereas this is an 
excellent approximation for the recoil-ion motion, it is less so for the 
neutrino.{\footnote{I owe this observation to Eli Friedman.}} Nevertheless, 
for a rest-frame isotropic distribution, it is estimated that neutrino forward 
angles in the laboratory dominate over backward angles by more than a factor 
five. 

For distinct diagonal Dirac-neutrino magnetic moments, Eq.~(\ref{eq:ratenu}) 
gets generalized to 
\begin{equation} 
{\cal R}_{\nu_e} = \lambda_{\rm EC}[1+2{\rm Re}(a_0 a^*_{\rm m}) 
\sum_j |U_{ej}|^2 \cos (\omega_{\mu_j}t)],  
\label{eq:ratenuj} 
\end{equation} 
resulting in a more involved pattern of modulation. 

For vanishing diagonal magnetic moments, and nonzero values of transition 
magnetic moments, the discussion proceeds identically to that for Majorana 
neutrinos in the next subsection.

\subsection{Majorana neutrino magnetic moments} 
\label{subsec:majorana} 

Majorana neutrinos can have no diagonal electromagnetic moments, but are 
allowed to have nonzero {\it transition} moments connecting different 
mass-eigenstate neutrinos, or different flavor neutrinos. A static magnetic 
field perpendicular to the storage ring will induce spin-flavor precession 
\cite{SVa81}. However, the magnetic interaction effect is masked in this case 
by neutrino mass differences, such that the amplitudes 
$\cos (\omega_{\mu_{\nu}}t')$ and $\sin (\omega_{\mu_{\nu}}t')$ in 
Eqs.~(\ref{eq:Aj}) and (\ref{eq:AjR}) are replaced, to leading order in 
$\omega_{\mu_{\nu}}/\Omega_{\nu_e} << 1$, by 
\begin{equation} 
\cos (\omega_{\mu_{\nu}}t') \rightarrow \exp(-{\rm i}\Omega_{jj'}t'),\,\,\,\,\, 
\sin (\omega_{\mu_{\nu}}t') \rightarrow \frac{\omega_{\mu_{jj'}}}{\Omega_{jj'}}
\sin (\Omega_{jj'}t'),  
\label{eq:transition} 
\end{equation} 
where $\hbar \omega_{\mu_{jj'}} = \mu_{jj'}\gamma B$, and $\Omega_{jj'}$ is 
defined by Eq.~(\ref{eq:correct}). 
The period of any oscillation that might be induced by these amplitudes 
is of order $\Omega_{\nu_e}^{-1}\sim 7\times 10^{-5}$~s which is several 
orders of magnitude shorter than the time resolution scale of $\sim 0.5$~s 
in the GSI experiment \cite{LBW08}. Therefore, such oscillations will 
completely average out to zero over realistic detection periods.

\section{Discussion and summary} 
\label{sec:sum} 

In conclusion, it was shown that interference terms between different 
propagating mass-eigenstate neutrino amplitudes in two-body EC reactions 
on nuclei cancel out to zero when no neutrinos are detected. 
Coherence between propagating mass-eigenstate neutrinos is 
evident within each one of the amplitudes for detecting a given flavor 
neutrino. It is only when all final flavor rates are summed upon incoherently, 
as motivated by the different flavor measurements required, that cancelations 
occur and the overall rate becomes independent of time and is reliably 
calculable from standard weak interaction theory for massless neutrinos. 
The underlying logic here is that summing on all possible phase space for 
flavor neutrinos is equivalent within quantum mechanics to not detecting any 
specific neutrino. 

Interference terms from different propagating mass-eigenstate neutrinos 
would survive and give rise to oscillatory behavior of the EC decay rate, 
if and {\it only} if neutrinos are detected. It was shown that the period 
of oscillations in such a case is 
$T\sim 4\pi E_{\nu}/{\Delta (m_{\nu}^2)}$ which for 
$E_{\nu}\approx 4$~MeV as in the GSI experiments \cite{LBW08}, and for 
$\Delta (m_{\nu}^2) \approx 0.76 \times 10^{-4}$~eV$^2$ \cite{SNO08}, 
assumes the value $T\sim 4.4\times 10^{-4}$~s, shorter by over four orders 
of magnitude than the period reported in these experiments. The oscillation 
period cited here is in full agreement with the oscillation length tested 
in dedicated neutrino oscillation experiments, provided the time $t$ is 
identified with $L/c$ where $L$ is the distance traversed by the neutrino. 
In particular, besides the $\Delta (m_{\nu}^2)$ neutrino input, it depends 
on the neutrino energy $E_{\nu}$, not on the mass $M_D$ of the decay ion. 

On the positive side, I have proposed a possible explanation of 
the `GSI Oscillations' puzzle connected with the magnetic field that guides 
the heavy-ion motion in the ESR, requiring a Dirac neutrino magnetic moment 
$\mu_{\nu}$ about six times smaller than the laboratory upper limit value 
from the Borexino Collaboration \cite{Borexino08}. The underlying mechanism 
is the interference between decay amplitudes not affected by the static 
magnetic field of the ESR and decay amplitudes affected by this field which 
induces spin precession of the emitted neutrino. The motion of the recoil ion 
in the ESR is constrained by the interference long after the neutrino has 
fled away. This mechanism does not work for Majorana neutrinos that can have 
no {\it diagonal} magnetic moments. For nonzero values of {\it transition} 
magnetic moments, the resulting spin-flavor precession is suppressed by 
neutrino mass differences, and it becomes impossible to relate then the GSI 
Oscillations puzzle to magnetic effects. It is not yet resolved experimentally 
whether neutrinos are Dirac or Majorana fermions, although the theoretical 
bias rests with Majorana fermions, in which case the present paper 
accomplished nothing towards providing a credible explanation of this puzzle. 

For experimental verification, note that the time-modulation period 
$T^{\rm lab}_{\rm EC}$ is inversely proportional to $B$, so the effect 
proposed here may be checked by varying $B$, for example by varying 
$\beta=v/c$ for the coasting decay ions. For a fixed value of $\beta$, 
$B$ depends on the charge-to-mass ratio of the decay ion which varies 
only to a few percent with the decay-ion mass $M_D$. Finally, the proposed 
effect is unique to two-body EC reactions, since three-body weak decays do 
not constrain the neutrino direction of motion with respect to the fixed 
direction of $\vec B$. Indeed, preliminary data on the three-body $\beta^+$ 
decay of $^{142}$Pm indicate no time modulation of the $\beta^+$ decay rate, 
limiting its modulation amplitude to $a_{\beta^+} < 0.03(3)$ \cite{Kienle09}.

\section*{Acknowledgement} Critical comments by Thomas Faestermann, Eli 
Friedman and Koichi Yazaki, and stimulating discussions with Paul Kienle 
and Harry Lipkin, are greatly appreciated. The support by the DFG Cluster 
of Excellence `Origin and Structure of the Universe' during a three-month 
visit in 2009 to the Technische Universit\"{a}t M\"{u}nchen, where 
Sect.~\ref{sec:magfield} was conceived, is gratefully acknowledged.

\end{document}